# Anharmonic phonon-phonon scattering at interface by non-equilibrium Green's function formalism

Yangyu Guo [1*], Zhongwei Zhang [1], Marc Bescond [2], Shiyun Xiong [1], Masahiro Nomura [1†], Sebastian Volz [1, 2‡]

[1] *Institute of Industrial Science, The University of Tokyo, Tokyo 153-8505, Japan*
[2] *LIMMS, CNRS-IIS UMI 2820, The University of Tokyo, Tokyo 153-8505, Japan*

(Dated: Mar. 14, 2021)

## Abstract

The understanding and modeling of inelastic scattering of thermal phonons at a solid/solid interface remain an open question. We present a fully quantum theoretical scheme to quantify the effect of anharmonic phonon-phonon scattering at an interface via non-equilibrium Green's function (NEGF) formalism. Based on the real-space scattering rate matrix, a decomposition of the interfacial spectral energy exchange is made into contributions from local and non-local anharmonic interactions, of which the former is shown to be predominant for high-frequency phonons whereas both are important for low-frequency phonons. The anharmonic decay of interfacial phonon modes is revealed to play a crucial role in bridging the bulk modes across the interface. The overall quantitative contribution of anharmonicity to thermal boundary conductance is found to be moderate. The present work promotes a deeper understanding of heat transport at the interface and an intuitive interpretation of anharmonic phonon NEGF formalism.

---

[*] yyguo@iis.u-tokyo.ac.jp
[†] nomura@iis.u-tokyo.ac.jp
[‡] volz@iis.u-tokyo.ac.jp

Heat transport at solid/solid interface is a critical issue in modern technologies and engineering applications such as thermal management of micro- and nano-electronics [1], nanostructured thermoelectrics [2], quantum cascade laser [3], phase-change memory [4], and so on [5]. However, a full understanding and modeling of interface conductance (or thermal boundary conductance, TBC hereafter) remain still an open question due to the broken translational symmetry of crystal lattice and complicated interface conditions [5,6].

TBC is currently described by two prevailing theories, i.e. the acoustic mismatch and the diffuse mismatch models (AMM [7] and DMM [8] respectively). The AMM, based on an elastic wave picture of phonons specularly transmitted across the interface, is usually valid at very low temperatures [8-10]. In contrast, the DMM assumes phonons as particles diffusely transmitted across an interface and captures the general trend of TBC at elevated temperatures. However, only limited agreement between the DMM and experimental data is achieved [6,8,11] since (*i*) elastic scattering only is involved in the DMM despite few efforts to include the inelastic correction [12], (*ii*) accounting for the interface atomic structure details remains a challenging task but significantly influences the TBC.

Atomistic simulation methods provide a direct avenue to consider both the inelastic effect from anharmonicity and the interface atomic structure. Important progress has been made in the spectral decomposition of the TBC into elastic and inelastic contributions via molecular dynamics (MD) simulations [13-15]. A formalism has also been developed for the modal decomposition of the TBC [16,17], yet based on a non-canonical definition of the eigen-modes of the interface system. In contrast to the classical MD simulation, the non-equilibrium Green's function (NEGF) formalism [18-20] is a fully quantum approach allowing for direct input of the first-principle atomic interaction force constants. However, ballistic NEGFs have been mostly adopted so far for the prediction of TBC [19,21-24] because of the computational challenge when including anharmonicity. Some attempts to include the anharmonicity at the interface based on an empirical probe approach have also been proposed [25,26]. In a recent contribution [27], an anharmonic NEGF formalism was developed to model the TBC and to demonstrate the significant role of anharmonicity at the interface. To sum up, the previous atomistic simulations generally provide the heat flow spectrum across the interface, while a clear understanding of how the anharmonicity involves in the alteration of the spectrum via phonon-phonon scattering process is still imperative. The emerged interfacial phonon modes have been shown to be important in interface heat transport [13,17,28,29]; however, it remains a mystery how the interfacial modes play the role. The contribution of anharmonicity to the TBC is also inconclusive due to the rigorous quantitative validation of anharmonic NEGF formalisms [30].

In this Letter, we present a theoretical model to extract and decompose the spectral energy exchange due to phonon-phonon scattering at an interface from the anharmonic

NEGF formalism developed in our recent work [30]. As a result, we demonstrate a clear understanding of how lattice anharmonicity contributes to phonon mode conversion and energy exchange at the interface. Especially we provide direct evidence of the anharmonic decay of interfacial phonon modes, which plays a crucial role in bridging the bulk modes from two sides. A quantitative contribution of anharmonicity to TBC is evaluated and is shown to be smaller than the one of the previous NEGF result.

We model heat transport across an ideally smooth Si/Ge interface as shown in Fig. 1. For simplicity, we assume that Si and Ge have the same lattice and force constants, with an only atomic mass difference, following the argument in Ref. [27,30]. The second- and third-order force constants are computed by first-principle method, with the details given in Ref. [30]. An interface region of a single-unit-cell in length is modeled considering that we focus on the transport mechanisms exactly around the interface.

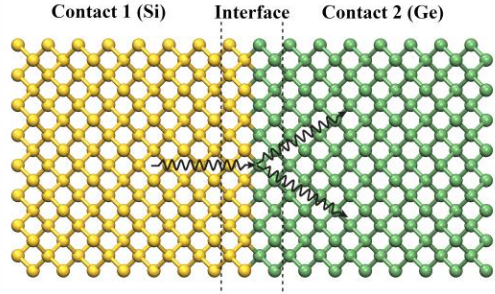

Fig. 1. Schematic of an anharmonic phonon-phonon scattering event at an ideal Si/Ge interface in the frame of non-equilibrium Green's function (NEGF) formalism.

The retarded Green's function of the interface region is computed in matrix notation as [30-33]:

$$\mathbf{G}^{R}(\omega;\mathbf{q}_{\perp}) = \left[\omega^{2}\mathbf{I} - \tilde{\mathbf{\Phi}}(\mathbf{q}_{\perp}) - \mathbf{\Sigma}^{R}(\omega;\mathbf{q}_{\perp})\right]^{-1}, \qquad (1)$$

where $\mathbf{I}$ is the unity matrix and $(\omega;\mathbf{q}_{\perp})$ denotes the frequency and wave-vector dependences along the transport and transverse directions, respectively. The retarded self-energy matrix includes the contribution from the two contacts and the anharmonic interaction in the interface region, i.e. $\mathbf{\Sigma}^{R}(\omega) = \mathbf{\Sigma}_{1}^{R}(\omega) + \mathbf{\Sigma}_{2}^{R}(\omega) + \mathbf{\Sigma}_{s}^{R}(\omega)$. The greater/lesser Green's function of the interface region is computed as [30-33]:

$$\mathbf{G}^{>,<}(\omega;\mathbf{q}_{\perp}) = \mathbf{G}^{R}(\omega;\mathbf{q}_{\perp})\mathbf{\Sigma}^{>,<}(\omega;\mathbf{q}_{\perp})\mathbf{G}^{A}(\omega;\mathbf{q}_{\perp}), \qquad (2)$$

where the advanced Green's function $\mathbf{G}^{A}(\omega;\mathbf{q}_{\perp})$ is the Hermitian conjugate of the retarded one. The greater/lesser self-energy matrix also includes the contribution from the contacts and the anharmonic interaction: $\mathbf{\Sigma}^{>,<}(\omega) = \mathbf{\Sigma}_{1}^{>,<}(\omega) + \mathbf{\Sigma}_{2}^{>,<}(\omega) + \mathbf{\Sigma}_{s}^{>,<}(\omega)$. The retarded scattering self-energy matrix is computed as [30,32,33]:

$$\Sigma_s^R(\omega;\mathbf{q}_\perp) = \frac{1}{2}\left[\Sigma_s^>(\omega;\mathbf{q}_\perp) - \Sigma_s^<(\omega;\mathbf{q}_\perp)\right] + i\mathrm{P}\int_{-\infty}^{\infty}\frac{d\omega'}{2\pi}\frac{\Sigma_s^>(\omega';\mathbf{q}_\perp) - \Sigma_s^<(\omega';\mathbf{q}_\perp)}{\omega - \omega'}, \quad (3)$$

where P denotes the Cauchy principal part of the integral. The greater/lesser scattering self-energy matrix is computed as [30]:

$$\begin{aligned}\Sigma_{s,l_x l_x'}^{>,<\,ij}(\omega;\mathbf{q}_\perp) = \frac{1}{2}i\hbar \sum_{l_{1x}l_{2x}l_{3x}l_{4x}} \sum_{j_1 j_2 j_3 j_4} \frac{1}{N}\sum_{\mathbf{q}_\perp'} \tilde{\Phi}_{l_x l_{1x} l_{2x}}^{ij_1 j_2}(\mathbf{q}_\perp',\mathbf{q}_\perp - \mathbf{q}_\perp') \tilde{\Phi}_{l_x' l_{3x} l_{4x}}^{jj_3 j_4}(\mathbf{q}_\perp' - \mathbf{q}_\perp, -\mathbf{q}_\perp') \\ \times \int_{-\infty}^{\infty}\frac{d\omega'}{2\pi} G_{l_{1x}l_{4x}}^{>,<\,j_1 j_4}(\omega';\mathbf{q}_\perp') G_{l_{2x}l_{3x}}^{>,<\,j_2 j_3}(\omega - \omega';\mathbf{q}_\perp - \mathbf{q}_\perp')\end{aligned}, \quad (4)$$

where the subscripts $l_x$ ($l_{1x}$, $l_{2x}$, …) denote the atomic index, and the superscripts $i$, $j$ ($j_1$, $j_2$, …) denote the cartesian coordinates ($x$, $y$, $z$), $N$ being the number of transverse wave vectors. The calculation of the contact self-energy matrices and the Fourier's representation of harmonic and anharmonic force constant matrices ($\tilde{\Phi}(\mathbf{q}_\perp)$ in Eq. (1) and $\tilde{\Phi}_{l_x l_{1x} l_{2x}}^{ij_1 j_2}(\mathbf{q}_\perp,\mathbf{q}_\perp')$ in Eq. (4)) can be found in our previous work [30]. The numerical solutions of the Green's function and self-energy matrices in Eqs. (1)-(4) are obtained by a self-consistent Born approximation iteration process [30,33].

In terms of physical interpretation, $i\mathbf{G}^<$ and $i\mathbf{G}^>$ denote the matrix generalization of the phonon occupation number in the present state ($n$) and that in the final state after the in-scattering process ($1+n$) in the Boltzmann transport theory [34], respectively, whereas $i\Sigma^<$ and $i\Sigma^>$ denote the matrix generalization of the in- and out-scattering rates separately. Similar arguments can be found in electron NEGF [35] except a sign difference due to the different statistics of electrons and phonons (fermions versus bosons). Therefore, the net difference of energy flux between out-scattering and in-scattering in the interface region due to contacts yields the heat flux formulas which were differently derived [32,36]:

$$\mathrm{J}_1 = \int_0^\infty d\omega \mathrm{J}_{1\omega}, \quad \mathrm{J}_2 = \int_0^\infty d\omega \mathrm{J}_{2\omega}, \quad (5)$$

where $\mathrm{J}_1$ and $\mathrm{J}_2$ denote the heat flux from contact 1 to interface region, and from interface region to contact 2, respectively, with the following full expressions of spectral heat fluxes:

$$\mathrm{J}_{1\omega} = \frac{\hbar\omega}{2\pi}\frac{1}{A_c N}\sum_{\mathbf{q}_\perp}\mathrm{Tr}\left[\Sigma_1^>(\omega;\mathbf{q}_\perp)\mathbf{G}^<(\omega;\mathbf{q}_\perp) - \Sigma_1^<(\omega;\mathbf{q}_\perp)\mathbf{G}^>(\omega;\mathbf{q}_\perp)\right], \quad (6)$$

$$\mathrm{J}_{2\omega} = \frac{\hbar\omega}{2\pi}\frac{1}{A_c N}\sum_{\mathbf{q}_\perp}\mathrm{Tr}\left[\Sigma_2^<(\omega;\mathbf{q}_\perp)\mathbf{G}^>(\omega;\mathbf{q}_\perp) - \Sigma_2^>(\omega;\mathbf{q}_\perp)\mathbf{G}^<(\omega;\mathbf{q}_\perp)\right], \quad (7)$$

where 'Tr' denotes the trace of a square matrix and $A_c$ is the transverse cross-sectional area. Eqs. (6) and (7) are valid for both ballistic and interacting phonon transport [37]. When the anharmonic interaction is considered as in the present work, similarly to the electron NEGF [38,39], the scattering self-energy shall satisfy the following energy conservation condition:

$$\delta \mathrm{J} = \int_0^\infty d\omega \delta \mathrm{J}_\omega = 0, \quad (8)$$

where $\delta J$ denotes the overall energy exchange due to anharmonic phonon-phonon scattering with $\delta J_\omega$ being its spectral component:

$$\delta J_\omega = \frac{\hbar\omega}{2\pi}\frac{1}{A_c N}\sum_{\mathbf{q}_\perp}\text{Tr}\left[\mathbf{\Sigma}_s^>(\omega;\mathbf{q}_\perp)\mathbf{G}^<(\omega;\mathbf{q}_\perp) - \mathbf{\Sigma}_s^<(\omega;\mathbf{q}_\perp)\mathbf{G}^>(\omega;\mathbf{q}_\perp)\right]. \quad (9)$$

In Eq. (9), the first term ($\mathbf{\Sigma}_s^>\mathbf{G}^<$) and second term ($\mathbf{\Sigma}_s^<\mathbf{G}^>$) represent the out- and in-scattering phonon numbers separately (except a factor of negative sign due to the imaginary unit: $i^2$). Thus $\delta J_\omega > 0$ and $\delta J_\omega < 0$ denote respectively net phonon generation and annihilation at a specific frequency $\omega$. In the ballistic limit, the scattering self-energy vanishes, which gives $\delta J_\omega = 0$. As a result, Eq. (9) provides a quantitative evaluation of the contribution of anharmonic phonon-phonon scattering to the mode conversion and energy exchange in the interface region.

As a further step, the overall spectral energy exchange $\delta J_\omega$ in Eq. (9) is decomposed into the contribution from different atom sites in the interface region: $\delta J_\omega = \sum_n (\delta J_\omega)_n$, with:

$$(\delta J_\omega)_n = \frac{\hbar\omega}{2\pi}\frac{1}{A_c N}\sum_{\mathbf{q}_\perp}\sum_{m,ij}\left[\Sigma_{s,nm}^{>,ij}(\omega;\mathbf{q}_\perp)G_{mn}^{<,ji}(\omega;\mathbf{q}_\perp) - \Sigma_{s,nm}^{<,ij}(\omega;\mathbf{q}_\perp)G_{mn}^{>,ji}(\omega;\mathbf{q}_\perp)\right]. \quad (10)$$

The on-site spectral energy exchange $(\delta J_\omega)_n$ includes the contribution from both local scattering self-energy when $m = n$ and non-local ones when $m \neq n$. As the translational invariance is broken along the transport direction around the interface, the conventional concept of modal scattering rate (inverse of a lifetime) in Boltzmann transport theory [34] becomes no longer valid. For interface heat transport, the real-space scattering rate matrix in Eq. (4) is more relevant and useful for evaluating the effect of anharmonicity. As indicated in Eq. (10), the diagonal ($m = n$) and off-diagonal ($m \neq n$) blocks of this matrix represent the contribution from the local and non-local anharmonic interactions, respectively. Therefore, it is natural to further decompose the on-site spectral energy exchange as: $(\delta J_\omega)_n = \sum_m (\delta J_\omega)_{nm}$, with:

$$(\delta J_\omega)_{nm} = \frac{\hbar\omega}{2\pi}\frac{1}{A_c N}\sum_{\mathbf{q}_\perp}\sum_{ij}\left[\Sigma_{s,nm}^{>,ij}(\omega;\mathbf{q}_\perp)G_{mn}^{<,ji}(\omega;\mathbf{q}_\perp) - \Sigma_{s,nm}^{<,ij}(\omega;\mathbf{q}_\perp)G_{mn}^{>,ji}(\omega;\mathbf{q}_\perp)\right], \quad (11)$$

or written in matrix notation as:

$$(\delta J_\omega)_{nm} = \frac{\hbar\omega}{2\pi}\frac{1}{A_c N}\sum_{\mathbf{q}_\perp}\text{Tr}\left[\mathbf{\Sigma}_{s,nm}^>(\omega;\mathbf{q}_\perp)\mathbf{G}_{mn}^<(\omega;\mathbf{q}_\perp) - \mathbf{\Sigma}_{s,nm}^<(\omega;\mathbf{q}_\perp)\mathbf{G}_{mn}^>(\omega;\mathbf{q}_\perp)\right]. \quad (12)$$

For the convenience of analysis, we introduce the layer-dependent on-site spectral energy exchange and decompose it into local and non-local contributions as:

$$(\delta J_\omega)_I = \sum_{n \propto I}(\delta J_\omega)_n = \sum_{J=I}(\delta J_\omega)_{I,J} + \sum_{J \neq I}(\delta J_\omega)_{I,J}, \quad (13)$$

where $(\delta J_\omega)_{I,J} = \sum_{n \propto I, m \propto J}(\delta J_\omega)_{nm}$ and the subscripts '$I$, $J$' denote the index of atomic layers in the interface region ($1 \leq I, J \leq 4$ here).

We conduct NEGF simulations of the heat transport across the Si/Ge interface from 50K to 600K. A small temperature difference (4K for T < 200K and 10K otherwise) is applied to ensure that the heat transport remains in the linear regime. A mesh of $N_\omega$=151 and $N = 9 \times 9$ for frequency and transverse wave vector points are adopted for all the cases after careful independence verification.

Firstly, we discuss the numerical results and theoretical analysis at 500K since the anharmonic interaction is significant at elevated temperatures. The spectral heat flux from Si contact to the interface region ($J_{1\omega}$) and from the interface region to Ge contact ($J_{2\omega}$) are shown in Fig. 2(a). In contrast to the elastic harmonic limit, an appreciable contribution to the heat flux arises from Si phonons beyond the cut-off frequency of Ge phonons. It is attributed to the anharmonic phonon-phonon scattering in the interface region, which is quantitatively described by the overall spectral energy exchange $\delta J_\omega$ in Eq. (9) as shown in Fig. 2(b). Very strong phonon annihilation ($\delta J_\omega < 0$) and phonon generation ($\delta J_\omega > 0$) are observed respectively in the high-frequency range (10~15 THz) and in the moderate-frequency range (5~9 THz), which exactly corresponds to the range of enhancement in heat flux spectrum ($J_{1\omega}$ and $J_{2\omega}$ separately) in Fig. 2(a). Thus we obtain an overall picture of how anharmonic scattering plays a role in the interface region: the high-frequency phonons incident from the Si contact are annihilated and the moderate-frequency phonons are generated, then leaving towards the Ge contact. A detailed energy balance relation is valid as demonstrated in Fig. 2(b): $\delta J_\omega = J_{2\omega} - J_{1\omega}$, which can be deduced from their definitions in Eqs. (6), (7) and (9) with the help of a universal relation in the NEGF formalism [38]: $\mathrm{Tr}(\mathbf{\Sigma}^<\mathbf{G}^> - \mathbf{\Sigma}^>\mathbf{G}^<) = 0$. Furthermore, the layer-dependent on-site spectral energy exchange computed from Eq. (10) and Eq. (13) in Fig. 2(b) quantifies the effect of anharmonic scattering in each layer from the Si contact towards the Ge contact. Strong phonon annihilation in the intermediate two layers (layers 2 and 3) is seen in the frequency range around 12 THz, which corresponds to the interfacial phonon modes as inferred from the layer-dependent local density of states (LDOS) in Fig. 2(c). The spatial distribution of LDOS indicates that the interfacial modes only exist within 1~2 layers away from the exact smooth interface, which is consistent with previous MD simulations of the Si/Ge interface [28,29]. Phonon annihilation around 14THz in the first two Si layers (layers 1 and 2) can be interpreted by the presence of optical modes. In the moderate-frequency range, phonon generation in all the four layers is clearly appearing. Therefore the decay of interfacial modes is crucial in shifting the energy of high-frequency phonons from the Si side to that of moderate-frequency phonons at the Ge side. Besides, there is even considerable spectral energy exchange in the lower frequency range (2~5 THz) in all layers although they compensate each other to some extent. These phonons shall involve in the three-phonon scattering processes with higher-frequency phonons. The phonons in the low-frequency limit (0~2THz) basically transmit through the interface in an elastic way, which is relevant at low temperatures.

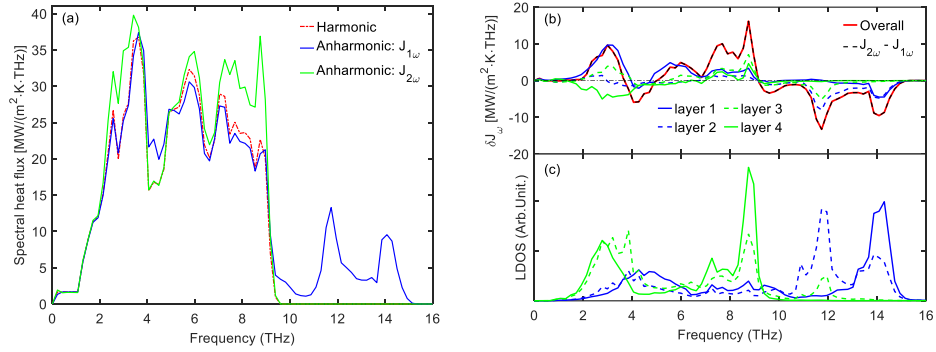

Fig. 2. Heat transport across the Si/Ge interface at 500K: (a) spectral heat flux per unit temperature difference, $J_{1\omega}$ and $J_{2\omega}$ denote heat flux from the Si contact to the interface, and from the interface to the Ge contact, respectively; (b) spectral energy exchange per unit temperature difference due to anharmonic phonon-phonon scattering in each layer of the interface region from the Si contact towards the Ge contact, the solid red line denotes the overall result in the interface region, and the dash-dot line is a reference of the harmonic limit; (c) local density of states (LDOS) in each layer of the interface region.

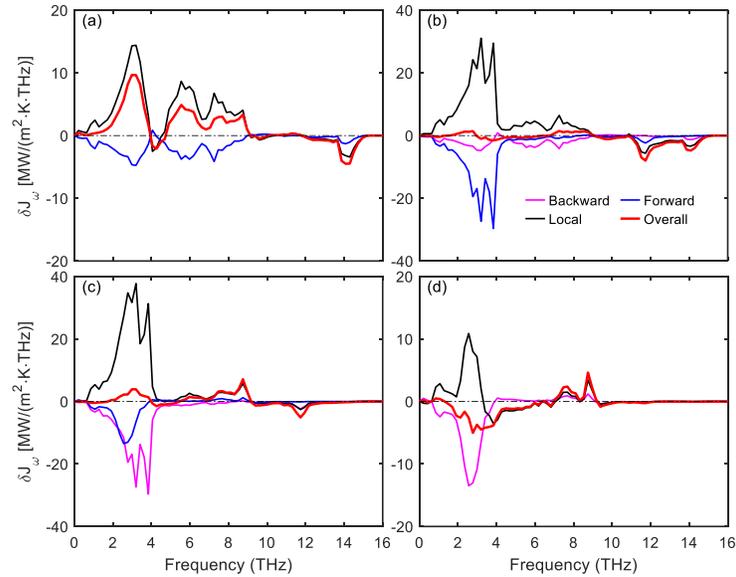

Fig. 3. Decomposition of the spectral energy exchange per unit temperature difference due to anharmonic phonon-phonon scattering in the four layers of the interface region at 500K: (a) layer 1; (b) layer 2; (c) layer 3; (d) layer 4 from the Si contact towards the Ge contact. The magenta line, black line and blue line represent the contribution from backward, local and forward scattering, respectively, whereas the solid red line denotes the overall result in each layer. The dash-dot line is the reference of the harmonic limit.

To gain a deeper understanding, we decompose the layer-dependent on-site spectral energy exchange into the local and non-local contributions based on Eq. (11) and Eq. (13), as shown in Fig. 3. In the non-local contribution, the backward term $(\delta J_\omega)_{I, I-1}$ and forward term $(\delta J_\omega)_{I, I+1}$ are merely considered because further terms of the scattering self-energy are negligibly small and not accounted in our anharmonic NEGF framework [30]. Note that the backward and forward terms for the first and last atomic layer respectively do not appear. In general, the local contribution $(\delta J_\omega)_{I, I}$ becomes predominant at moderate-to-high frequency (> 6~8 THz) and is quite close to the overall spectral energy exchange in each of the four layers. This indicates that the anharmonic scattering of high-frequency phonons at the interface is very local from the real-space point of view. It makes sense as the high-frequency phonons of Si (or Ge) usually have extremely short wavelengths close to atomic separation. In contrast, the non-local contribution is very large at low frequency (< 5THz), especially in the intermediate two layers shown in Fig. 3(b) and Fig. 3(c), although it much counteracts the local contribution finally. Physically speaking, both the local and non-local real-space anharmonic scattering are important at the interface for low-frequency phonons usually with longer wavelengths. From the modeling perspective, both the diagonal and off-diagonal blocks in the scattering self-energy matrix in Eq. (4) are indispensable. The previous anharmonic phonon NEGF formalism considering only the local scattering self-energy [33] would fail to capture the behaviors of those low-frequency phonons accurately. As shown in Fig. 2(a), these phonons (2~4THz) have a non-negligible anharmonic contribution to interface heat flux. In addition, the forward and backward terms in the non-local contribution are found to be reciprocal between neighboring layers: $(\delta J_\omega)_{I, I+1} = (\delta J_\omega)_{I+1, I}$. This can be verified rigorously using symmetrical relations between Green's function and self-energy matrices, as described in the Supplementary Materials [40]. The decomposition of the layer-dependent on-site spectral energy exchange displays a similar trend at other elevated temperatures, as shown in Fig. S1 at 300K and Fig. S2 at 600K [40].

Finally, the temperature dependence of the interface heat transport is discussed. With increasing temperature, the spectral heat flux from the Si contact to the interface region has increasing enhancement beyond the cut-off frequency of Ge, as shown in Fig. 4(a). In the low-temperature limit as in the case at 50K, the anharmonic NEGF result almost coincides with the harmonic one since the phonon-phonon scattering is very weak. The increasing trend of the enhancement can be understood from the temperature dependence of the overall spectral energy exchange in the interface region due to anharmonicity as reported in Fig. 4(b). The amplitude of the two dips in the high-frequency range (>10 THz) gradually rises with temperature due to respectively the decay of interfacial phonon modes and optical phonon modes according to our preceding discussion. The growing spectral energy exchange is mainly caused by the increase of phonon occupation number of higher-frequency modes, which in turn strengthens the real-space anharmonic scattering as indicated by Eq. (4). As a result, the TBC increases with temperature as demonstrated in Fig. 4(c), where the difference between the anharmonic result and the harmonic one also becomes larger at higher temperature. The TBC is enhanced due to the anharmonic phonon-phonon scattering at the interface, as is further corroborated by the ratio of anharmonic to harmonic TBCs in Fig. 4(d). The enhancement of the TBC is about 10% at room temperature and reaches about 20% at 600K. Those figures are appreciably smaller than the results in a very recent study of the same problem via anharmonic phonon NEGF as shown in the insets of Fig. 4(c) and Fig. 4(d) [27]. Note that in our previous work [30], a rigorous quantitative validation was proposed of our anharmonic phonon NEGF formalism, which ensures both the energy conservation in Ref. [27] and quasi-momentum conservation in the phonon-phonon scattering events. The present result is more or less consistent with the conclusions of previous MD simulations [28,41,42], i.e. the effect of anharmonicity away from the interface is more significant than that exactly at the interface. As there is no robust experimental data of the TBC for the Si/Ge interface, further work is pending to make a direct comparison to experimental results of more realistic interfaces with strong benchmark data [11]. Nevertheless, the present anharmonic phonon NEGF formalism and theoretical model are universally applicable to other solid/solid interfaces.

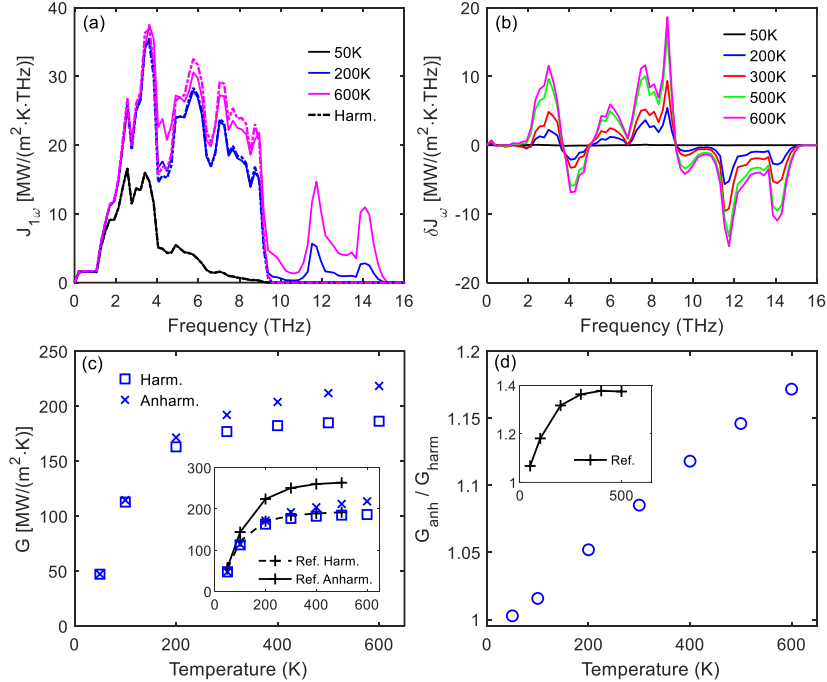

Fig. 4. Temperature dependence of heat transport across Si/Ge interface: (a) spectral heat flux per unit temperature difference from Si to interface region, the solid lines and dash-dot lines denote the anharmonic and harmonic results, respectively; (b) the overall spectral energy exchange per unit

temperature difference in the interface region due to anharmonic phonon-phonon scattering; (c) thermal boundary conductance, the square and cross symbols denote the harmonic and anharmonic results separately; (d) thermal boundary conductance ratio. The plus symbol with the line in the inset of (c) and (d) denote the reference result from Ref. [27].

In summary, a theoretical scheme is presented for quantification and decomposition of the spectral energy exchange due to phonon-phonon scattering at interface via a NEGF formalism. We promote the concept of real-space anharmonic phonon scattering rate for heat transport across an interface system with broken symmetry. The local interaction is shown to dominate the anharmonic scattering of high-frequency phonons, whereas both local and non-local interactions are significant for that of low-frequency phonons. Direct evidence is demonstrated of the decay of interfacial modes at the interface, which plays a crucial role in bridging the bulk modes away from the interface. The overall contribution of anharmonicity at the interface to thermal boundary conductance is found to be moderate. This work provides a deeper exposition of the physics of interface heat transport. The physical interpretation and theoretical analysis of the anharmonic phonon NEGF simulation will also advance a more intuitive understanding and its broader application.

**Acknowledgements**

This work was supported by the Postdoctoral Fellowship of Japan Society for the Promotion of Science (P19353), and CREST Japan Science and Technology Agency (JPMJCR19I1 and JPMJCR19Q3). This research used the computational resource of the Oakforest-PACS supercomputer system, The University of Tokyo.